%% file: chapter.tex
\documentclass[a4paper,11pt]{article}
\usepackage{aaskaiid}
\usepackage{soul}
\usepackage{orcidlink}

\title{{Overview: Extragalactic Continuum Science with the SKAO}}
\ShortTitle{SKAO Extragalactic Continuum SWG}

\author[1]{{Catherine L. Hale}$^{\orcidlink{0000-0001-6279-4772}}$}
\author[2,3,4]{{Fatemeh S. Tabatabaei}$^{\orcidlink{0000-0002-0377-0970}}$}

\affiliation[1]{{Institute for Astronomy, School of Physics and Astronomy, University of Edinburgh,  Royal Observatory Edinburgh, Blackford Hill, Edinburgh, EH9 3HJ, UK}}
\emailAdd{{Catherine.Hale@ed.ac.uk}}
\affiliation[2]{{School of Astronomy, Institute for Research in Fundamental Sciences (IPM), PO Box 19395-5531, Tehran, Iran}}
\affiliation[3]{I. Physik. Institut, University of Cologne, D-50937 Cologne, Germany}
\affiliation[4]{{Max-Planck-Institut f\"ur Astronomie, Department of Galaxies and Cosmology, K\"onigstuhl 17, D-69117 Heidelberg, Germany}}
\emailAdd{{ftaba@ipm.ir/tabata@ph1.uni-koeln.de}}

\abstract{Radio continuum observations provide a powerful probe of energetic processes that drive galaxy evolution across cosmic time. The {combined} sensitivity, angular resolution, and survey speed of the {SKAO's telescopes} will enable transformative advances in extragalactic astronomy by revealing {important details into} the interplay between star formation, accretion onto supermassive black holes, magnetic fields, and cosmic rays in galaxies and their environments. 
In this article, we summarize the key science goals of the {Extragalactic Continuum Science Working Group} {and the contributions of related chapters to Advancing Astrophysics with the SKA II (AASKAII)}. {For star forming galaxies, this includes using} multi-band SKA continuum observations to provide dust-unbiased measurements of the cosmic
star-formation history and enable robust spectral energy distribution analyses of
star-forming galaxies (SFGs) across cosmic time.  
{This is alongside studies which will be crucial to uncover} the physics and duty cycles of active galactic nuclei (AGN), clarify the origin of radio emission in radio-quiet AGN, and probe the co-evolution of black holes and their host galaxies. {Moreover, the}  SKA’s sensitivity  will also reveal diffuse synchrotron emission in galaxy clusters and the cosmic web, tracing cosmic-ray acceleration {within} large-scale magnetic fields. {Extragalactic continuum studies with the SKA} will combine wide area continuum surveys, multi-band studies and imaging over those fields which have an abundance of data across the electromagnetic spectrum.
{The capabilities of the SKA telescopes will position it} as a cornerstone facility for addressing fundamental questions in galaxy formation and evolution.}

\begin{document}
\maketitle
\include{journal-names}

\section{Introduction}
Radio continuum observations provide {a unique viewpoint of the Universe, tracing some of the key physical processes which can help} shape galaxies and their environments. Unlike spectral-line emission that traces specific atomic or molecular transitions, radio continuum emission arises from {two} physical mechanisms{, namely synchrotron and free-free emission.} Therefore, radio continuum observations can provide astronomers a unique insight {into physical processes such as star formation, AGN activity and the feedback mechanisms from these processes, alongside details on the prevalence of magnetic fields and cosmic rays}. 

Since the establishment of the SKA science working groups, the {Extragalactic Continuum Science Working Group (SWG)} has played a role in providing a conduit for interaction between the SKA's science and technical teams with the astronomical community. Activities of this SWG include defining and prioritizing science use cases and testing their feasibility, proposing reference surveys, providing feedback and assessments to {features such as the baseline design, frequency coverage} and science data processing. In addition, members of the {Extragalactic Continuum SWG} {(}which constitute the largest SKA science community{)} have collaborated on observations with SKA precursors or pathfinders, {have engaged with} SKA science data challenges, and {performed simulations to understand the expected science capabilities and surveys possible with the the SKA.}

\begin{figure}
    \centering
    \includegraphics[width=0.94\linewidth]{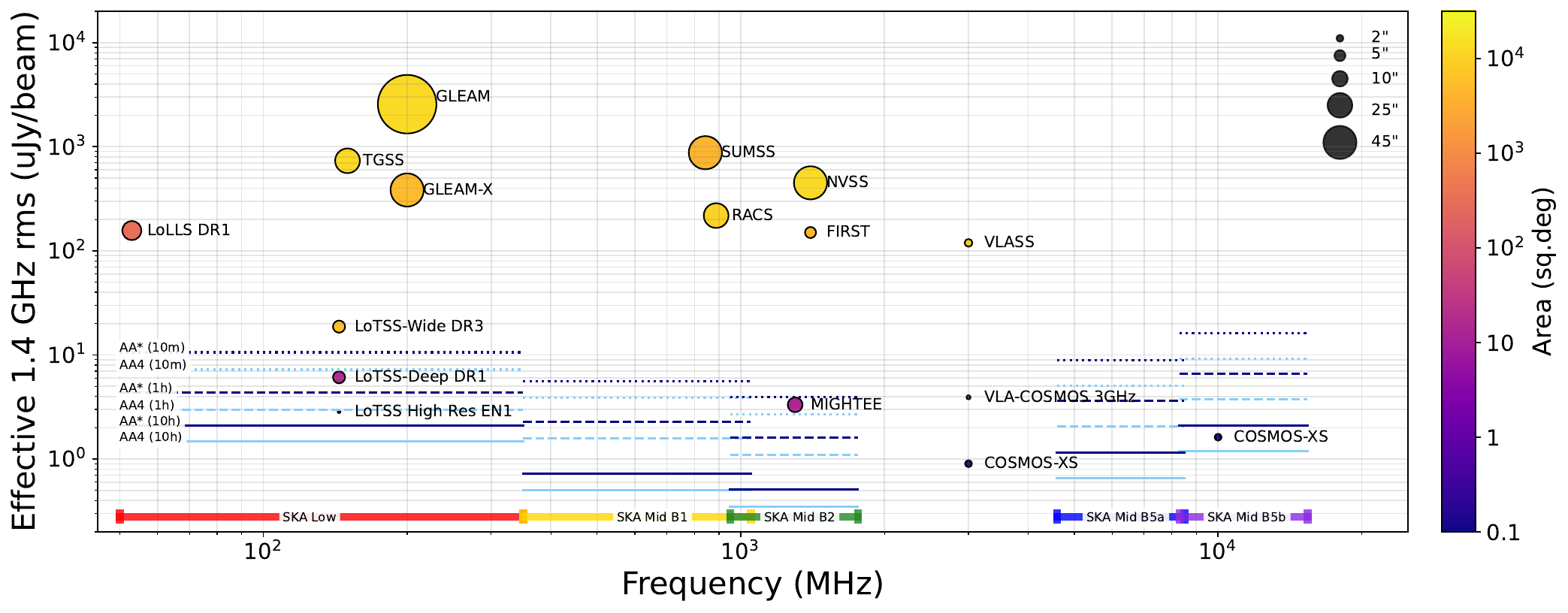} \vspace{-0.4cm}
    \caption{{{Equivalent 1.4 GHz sensitivity (assuming $S_{\nu} \propto \nu^{-0.7}$) for a subset of existing and ongoing surveys (coloured by sky area, size depicting resolution): LoLSS \citep[DR1][]{LoLSS}, GLEAM \citep{GLEAM}, GLEAM-X \citep[DR2][]{GLEAM-X}, TGSS \citep{TGSS}, LoTSS DR3 \citep{LoTSS-DR3}, LoTSS Deep DR1 \citep[over the multi-wavelenngth areas]{Sabater2021, Tasse2021, Kondapally2021}, LoTSS high resolution imaging over ELAIS-N1 \cite{deJong2024}, SUMSS \citep{SUMSS}, RACS-Low \citep{RACS}, EMU \citep[expectations, see][]{EMU}, NVSS \citep{NVSS}, FIRST \citep{FIRST}, MIGHTEE \citep{Hale2025a}, VLASS \citep[expectations, see][]{VLASS}, VLA 3GHz COSMOS \citep{VLA3GHz} and COSMOS-XS \citep{COSMOS-XS}. {Frequency bands from SKA-Low and SKA-Mid} are shown (coloured lines) with estimated sensitivities in 10m, 1h and 10h observations (dotted/dashed/solid lines) for AA* and AA4 (dark/light blue). These assume $\delta=$-30$^{\circ}$ and a Briggs' weighting of 0 (or -1 for SKA-Low, due to confusion).}}} 
    \label{fig:surveys}
\end{figure}

{Notably, a wealth of advances by SKA precursor and pathfinder facilities has transformed our views of the radio skies since the last SKA Book \citep[AASKAI;][]{bourke2015advancing}. This includes wide area surveys from telescopes at low frequencies ($<1$ GHz) including GLEAM\citep[-X;][]{GLEAM, GLEAM-X}, LoTSS \citep{LoTSS-DR3} and LoLSS \citep{LoLSS}. At high frequencies ($\gtrsim$1 GHz), these large area surveys are complemented through surveys including EMU \citep{EMU} and VLASS \citep{VLASS}. These surveys have provided vast numbers of radio sources {\citep[see e.g.][{where $\sim$14 million sources are detected}]{LoTSS-DR3}} {including the rarest and brightest sources,} whilst also providing {large} samples to trace a diverse range of environments and allowing for cosmological studies. Deeper surveys over smaller areas with an abundance of multi-wavelength data such as from the VLA 3GHz COSMOS survey \citep{VLA3GHz}, COSMOS-XS \citep{COSMOS-XS}, LOFAR Deep fields \citep[e.g.][]{Sabater2021} and MIGHTEE \citep{MIGHTEE} have provided {further observations which} have high levels of host galaxy associations {\citep[see e.g.][]{Algera_2020, Kondapally2021, Whittam2024}}. {These surveys allow} the evolution of AGN activity and the cosmic star formation history to be traced over large periods of time. A comparison of some of these surveys {with expected imaging capabilities from the SKA-Low and SKA-Mid} are shown in Figure \ref{fig:surveys}.}

{However, whilst having made striking advances in radio astronomy, deep surveys such as MIGHTEE are now limited by confusion in their images, hindering their ability to probe the key scientific questions across large periods of cosmic history \citep[see e.g.][]{Hale2025a}. Whilst this is being overcome at low frequencies using longer baselines \citep[see e.g.][]{Morabito2022}, {SKA-Mid} is essential for allowing deeper radio continuum studies of galaxy evolution {at $\sim$GHz frequencies}. Moreover, the combined spectral coverage of SKA-Low and SKA-Mid will be transformative in the modelling of the spectral energy distributions of radio sources across $\sim$3 orders of magnitude in frequency, from O(10 MHz-10 GHz).  Crucially, the {SKAO's telescopes are planned to have a significant improvement in survey speed} compared to current facilities. 
As presented in Figure \ref{fig:surveys} (see parameters used in the Figure), {$\sim$}1h of SKA-Low data at AA* will be comparable in depth to the LOFAR Deep Fields \citep{Sabater2021, Tasse2021} which used $\sim$100 hours per field. Similarly, just 10m with {SKA-Mid} (Band-2) at AA4 {is expected to} improve upon the depth of MIGHTEE, and comparable depths in Band 5 to the COSMOS-XS survey are possible within $\sim1/10$th of the time. Combined, increased survey areas can be covered within reasonable times (reducing cosmic variance limitations) and such sensitive observations can help drive our understanding of radio emission to more `ordinary' populations (e.g. lower $M_{*}$ and SFR) and trace populations to significantly higher redshifts. } 

Through our different science focus groups, the {Extragalactic Continuum} SWG explores {the extragalactic} Universe from nearby to high-redshift galaxies, examining structure formation and evolution on all scales. These focus groups include cosmic star formation, physics of AGNs, galaxy clusters, nearby galaxies, the interstellar and intergalactic medium (ISM \& IGM) at high redshifts, and gravitational lensing. In accordance with the aim of the AASKAII science book, these groups contribute to the advances in science with SKA through {33} chapters, {and their content reflect the key science goals which are a priority of the {Extragalactic Continuum} SWG, building on those discussed in the release of AASKAI}.  
{This includes a chapter by \cite{Prandoni01.2026.SKA}, which provides an updated discussion and outline of a proposed three-tiered reference surveys suggested for AASKAI \citep[in][]{Prandoni2015}, to probe galaxy evolution studies.} {In this chapter we review the} contributions {of the {Extragalactic}} Continuum SWG {and related science chapters} to the AASKAII science book in each of the science focus groups {within our SWG.}

\section{Star Formation History}
One of the primary goals of radio continuum observations is to trace dust-obscured star formation in the Universe. {Both the thermal free-free and the nonthermal synchrotron components of the radio continuum emission} in galaxies trace different evolutionary phases of young, massive stars \citep[see e.g.][]{Condon1992}. Together, these two components provide a dust-unbiased tracer of star formation in galaxies, both obscured and unobscured. As such, they can provide one of the most robust measurements of the star-formation history of the Universe \citep[see e.g.][]{Matthews2021, Cochrane2023}. {As SFGs are traditionally radio faint populations, dominating the extragalactic continuum populations at 1.4 GHz flux densities of $S \lesssim$ 0.2 mJy \citep[e.g.][]{Smolcic2017b}, the {SKA telescopes} will detect significant numbers of SFGs. Thus, {the surveys with the SKA} will provide large, unbiased populations to trace star formation to higher redshifts and beyond the peak of star formation \citep[$z\sim$2; see e.g.][]{Madau2014}. Expectations of SFGs from semi-empirical models for proposed surveys {with the SKA telescopes} are discussed in the chapter by \cite{Giulietti01.2026.SKA}.}

{However, to accurately probe cosmic star formation, understanding the relationships between radio luminosities and SFR is essential \citep[see e.g. studies in][]{Cook2024}. This calibration has traditionally relied on the infrared–radio correlation \citep[IRRC; see e.g.][]{Helou1985, Delvecchio2021}, however, the complexities of this relation are challenging.} {\cite{Taba17, taba_25} introduced more direct and robust SFR calibrations by analysing the radio spectral energy distribution (SED) in galaxies both nearby and at high redshift. The chapter by \cite{FangxiaAn01.2026.SKA} outlines the advantages of using the radio SEDs of galaxies in studies of star formation history.} 
{The chapter by} \cite{Algera01.2026.SKA} {echoes the importance of free-free emission in this role, and discusses how} such a reference survey will detect $\simeq 1.5 \times 10^4$ star-forming galaxies in all bands out to $z \simeq 7$, consistent with the findings of \cite{FangxiaAn01.2026.SKA} {and} \cite{Giulietti01.2026.SKA}.

The high sensitivity, resolution, field of view, and high survey speed of SKA AA4 will enable the construction of statistically robust samples of SFGs across cosmic time to trace the evolution of their radio luminosity functions {and the cosmic star formation history}. {Probing such high redshifts %will 
{may} introduce other challenges such as inverse Compton losses from the Cosmic Microwave Background \citep[CMB; see e.g.][]{Whittam2025} 
{and contamination from Anomalous Microwave Emission \citep[AME, see chapter of][]{Yoon01.2026.SKA}} could affect higher frequencies ($\gtrsim$10 GHz).}

\section{Active Galactic Nuclei}
AGN, powered by the accretion of matter onto supermassive black holes, are considered as fundamental drivers of galaxy evolution \citep[see e.g. reviews in][]{Fabian2012, Heckman2014, Harrison2024}. Through processes commonly described as AGN feeding and feedback, these systems regulate the growth of galaxies by redistributing energy and matter into the surrounding interstellar and circumgalactic media. Despite their importance, the physical mechanisms governing gas accretion onto black holes and the subsequent impact of AGN activity on star formation remain poorly understood {- the SKAO provides an important tool to address these}.

{Firstly,} AGN activity appears to be episodic but it is still unclear how long the AGN activity cycles last and what causes the switching on and off of radio jets. {Detailed studies of resolved AGN across broad frequencies {are} crucial to understand the mechanisms driving AGN activity, spectral ageing and} the duty cycle of AGN, {as} presented {in the chapter of \cite{Hardcastle01.2026.SKA}}. {Moreover,} how these short-lived AGN activity episodes connect to the much longer timescales of gas accretion and star formation is a central problem in modern extragalactic astrophysics. The importance of the SKA in resolving this problem is addressed through multi-band AA4 observations of continuum as well as HI line emission of hundreds of nearby AGNs, {see the chapter by \cite{Maccagni01.2026.SKA}.} 

{Beyond regulating and influencing the development of galaxies, understanding the interaction of AGNs within the surrounding environments is key to understanding the broader impact of feedback. This is addressed in a combination of chapters which outline topics such as: the morphologies of radio galaxies \citep[see chapter by][]{Sasmal01.2026.SKA} and how environments impact these morphologies \citep[see chapter by][]{SabyasachiPal03.2026.SKA}; how AGN feedback can impact galaxy groups {\citep[see the chapter of][]{Pasini01.2026.SKA} and intracluster medium through radio mini-halos \citep[see chapter of][]{Gitti01.2026.SKA}} and, finally, how magnetic fields can play a role in understanding the complex morphologies in radio galaxies \citep[see chapter by][]{Fromm01.2026.SKA}. Detailed studies using the combined frequency coverage of SKA-Low and Mid will help to uncover such complex interactions. }

{Given the importance of AGN jets in imparting energy and regulating galaxies through feedback, it is crucial to} understand how common radio jets were in the early universe and how they evolved across cosmic {time}. {Surveys with the SKA telescopes} can uniquely advance our knowledge about the AGN populations and their cosmic evolution, {probing significant numbers of AGN across luminosities and diverse environments to $z\sim6$} {\citep[see the chapter by][]{Kondapally01.2026.SKA}}. {The SKAO telescopes's sensitivity will also aid in in identifying the counterpart of AGNs {within the} epoch of reionization {\citep[$z\gtrsim 6$, see the chapter of][]{Afonso01.2026.SKA}} and also be a key tool to provide potential radio counterparts for sources such as AGN and `Little Red Dots' which have recently been detected by JWST {\citep[see chapter of][]{Mazzolari01.2026.SKA}.  }}
{Studies of compact AGN may also further benefit from the use of} interplanetary scintillation techniques {\citep[see chapter by][]{Chhetri01.2026.SKA}}.

{Finally, whilst existing surveys have often focused on the radio loud AGN, radio quiet AGN (RQAGN) {will} increase in numbers at the fainter flux densities accessible {with surveys from the SKA telescopes}.} The origin of the radio emission in RQAGN is still under debate {\citep[see e.g.][]{Panessa2019, Njeri2026}}. Possible mechanisms include weak compact jets, accretion disk coronae, AGN-driven winds or outflows, and host galaxy star formation. Observations with {SKA-Mid in AA4 will help understand the} origins and physics of radio continuum emission for RQAGN, {see the chapter by \cite{Kudoh01.2026.SKA} and {the chapter of \cite{Panessa01.2026.SKA} which makes use of VLBI}.}

\section{Galaxy Clusters and Large Scale Structure}

Galaxy clusters are the largest gravitationally bound systems in the Universe, and radio observations reveal their nonthermal components: relativistic particles and magnetic fields in the intracluster medium (ICM) - {see a review by \cite{vanWeeren2019}}. However, many fundamental questions remain, including the origin of radio halos, formation of radio relics, magnetic fields in the ICM, cosmic-ray transport and acceleration, and the evolution of cluster emission over cosmic time.

Radio halos are thought to be produced as a result of galaxy mergers. They are found {to be} bright at low frequencies with an ultra-steep spectrum, indicating that cosmic ray electrons are accelerated due to turbulence generated by cluster mergers. {The chapter of \cite{ArpanPal01.2026.SKA} discusses galaxy cluster mergers shocks and their detection at radio frequencies and how the combination of SKA-Low and X-ray data can help unearth the physics and evolution of such shocks.} Furthermore, as shown in {the chapter of \cite{Cassano01.2026.SKA}}, SKA-Low AA4 will probe an unprecedented region of cluster mass and redshift space, detecting at least ~2500 radio halos up to $z\simeq 0.6$, including about 1000 ultra-steep-spectrum systems.

Besides turbulence, cluster mergers also produce shocks which can compress magnetic fields, accelerate cosmic ray electrons, and create arc shaped radio relics on Mpc scales at the periphery of clusters. Similar structures known as {\it odd radio circles} (ORCs) have recently been found in galaxy groups on smaller scales of 150-500\,kpc which are thought to be circular shock fronts that occasionally occur around massive early-type galaxies during their evolution \citep[see chapters by][]{Koribalski01.2026.SKA, SabyasachiPal02.2026.SKA}. The unprecedented sensitivity and survey speed {from the SKA telescopes} will allow us to search for large numbers of these diffuse structures to study them in more detail. 
\vspace{-0.1cm}

{On the largest scales, the distribution of galaxies in the Universe forms an interconnected network known as the cosmic web \citep[see e.g. spectroscopic surveys such as {from}][]{DESI}.} This agrees with the standard $\Lambda$CDM cosmological paradigm, {where large-scale structure emerges} through the accretion of matter and hierarchical merging of smaller structures, ultimately leading to gravitational bound systems which are connected by a {cosmic web}. Similarly to clusters, {the cosmic web} must be filled with plasma but at lower temperature and density than clusters. Galaxy interactions and dynamical evolutions release energy and produce turbulence and shocks, which can amplify magnetic fields,  accelerate cosmic ray electrons and {produce synchrotron radiation (albeit much weaker than for galaxy clusters}.
As reviewed {in the chapter} by \cite{Cuciti01.2026.SKA}, SKA precursors and pathfinders have already detected components of the cosmic web (e.g. mega halos, bridges), but sensitive observations {with the SKA} will allow for fuller detection of structure from the cosmic web. In particular, wide SKA-Low observations will play a {role in detecting relatively low-energy plasma in the cosmic web}, and large and deep surveys of radio galaxies are proposed to trace cosmic webs at high redshifts {\citep[see the chapter of][]{Dabhade01.2026.SKA}}. {Furthermore,} whilst cosmic ray electrons are thought to be accelerated due to shocks or turbulence in cluster/cosmic web environments, the detailed microphysics and sources of shocks which contribute to such emission need further studies. Reviewing these, {the chapter of \cite{deGasperin01.2026.SKA}} highlights the SKA's fundamental importance in understanding acceleration and transport of these particles.

{As magnetic fields are crucial for nonthermal emission, studying} the origin and amplification mechanisms of magnetic fields in the ICM at high redshifts, {is key to understand the evolution of magnetic fields}. Evidence has been found for a rapid growth of the magnetic field within a few Gyrs of the Big Bang, challenging standard dynamo timescales \citep[see e.g.][]{Xu}. {The chapter of \cite{Santra01.2026.SKA}} {suggest studies of the massive merging system El Gordo with the SKA telescopes} {will help to assess such} models. 

Finally, the ICM also contains thermally ionized gas which can be studied in the radio through its interaction with CMB photons, known as the Sunyaev-Zeldovich (SZ) effect.  The SKA will be sensitive to the thermal SZ effect in its highest frequency Band\,5b {\citep[see the chapter by][]{Perrott01.2026.SKA}}. 

\vspace{-0.2cm}

\section{Detailed Astrophysics of Star Formation and Accretion in Local Galaxies}
{As outlined in this chapter,} the radio continuum emission from galaxies is dominated by star formation and AGN activity. {Whilst the combination of these processes can be challenging to disentangle \citep[but see][]{Morabito2025}, nearby} galaxies offer unique laboratories for disentangling AGN and star forming contributions to allow studies of their astrophysics in detail and at high resolutions. SKA-Low and Mid observations of nearby galaxies will ideally characterize different processes and radio sources through spectral analysis {\citep[see chapter of][]{Moldon01.2026.SKA}}, {making use of the SKA's broad frequency coverage}. {Moreover, {observations with the SKAO's telescopes} could help understand the nature of radio emission in ultra diffuse galaxies in the local Universe \citep[see the chapter of][]{Lal01.2026.SKA}.} 

\section{Strong Lensing}

Strong lensing with SKA will address several major unsolved problems in astrophysics and cosmology, including: the small-scale nature of dark matter; the internal mass structure and evolution of galaxies; {and the demographics of spiral, group, and cluster lenses} {\citep[see][]{McKean2015}}. 
The advantages of radio observations at high angular resolution, free from dust extinction, with broad spectral coverage over large areas will enable the discovery of large and less biased lens samples and support precision modeling of both lenses and lensed sources. In particular, {surveys using the SKA} will be powerful for detecting dark substructure, constraining galaxy mass profiles, and exploiting lensed radio AGN and SFGs as probes of both source physics and cosmic structure {\cite[see chapters by][]{McKean2015, Pandey-Pommier02.2026.SKA}. }

\vspace{-0.2cm}

\section{Role of ISM \& IGM Processes at High Redshifts in Evolution of Galaxies}
The ISM acts as the central engine of galaxy evolution, linking star formation, feedback, and gas flows. Gas inflows from the circumgalactic medium and the IGM replenish galaxies with fresh material for star formation, while outflows driven by supernovae or AGN can remove gas and regulate galaxy growth. The complex interplay between inflows, internal ISM processes, and feedback-driven outflows determines whether a galaxy continues forming stars or gradually becomes quiescent in the standard baryon cycle models of galaxy evolution \citep[see e.g.][]{Saintonge}. {However, the possible role of nonthermal pressures exerted by magnetic fields and cosmic rays is generally not incorporated into these scenarios \citep{taba_18,ghasemi}.}

{SKA observations} will enhance understanding of this interplay through continuum, HI 21-cm, and polarization observations, such as by mapping the thermal and nonthermal processes in galaxies up to cosmic noon and beyond {\citep[see chapter of][]{Tabatabaei01.2026.SKA}}. {These studies} will also address the role of external interactions in the galaxy quenching across different environments {through e.g. ram pressure stripping \citep[see chapter of][]{Ignesti01.2026.SKA}}.  

Deep, multi-band SKA surveys will further constrain the physical mechanisms driving the leakage of Lyman continuum photons, particularly the roles of supernova, radiative, and possibly cosmic-ray feedback in metal-poor, low-mass SFGs, which are considered important sources of cosmic ionizing photons {\citep[see chapter by][]{Bait01.2026.SKA}}. SKA-Mid Band 5 can also be used to explore anomalous microwave emission (AME) in distant galaxies {\citep[see chapter of][]{Yoon01.2026.SKA}}.

\section{Conclusions}
{Combined, the SKAO's telescopes and {their observations have the potential to derive} significant advances within the {scientific scope} of the {Extragalactic Continuum} SWG {by tracing} energetic processes over vast periods of cosmic time. This includes {providing} unique insights into how the processes of star formation and AGN accretion {shape} galaxy evolution{, as well as how the interplay between magnetic fields and cosmic rays probes diverse environments throughout the Universe.} These studies will rely on accurate source finding and characterisation \citep[see chapter by][]{SabyasachiPal01.2026.SKA} and will be aided by multi-wavelength host associations \citep[see e.g.][]{Duncan01.2026.SKA}. {Achieving these goals} will require a combination of statistical, machine learning and visual inspection techniques - the latter of which can {further benefit from engagement} beyond the SKAO community \citep[see chapter by][]{Hota01.2026.SKA}. Beyond the {Extragalactic Continuum SWG}, there is a wealth of synergies with other SWGs, as outlined in a number of other chapters throughout AASKAII {and at other wavelengths, such as with gamma rays \citep[see discussion in][]{Castignani01.2026.SKA}.}}

\section*{Acknowledgements}
CLH acknowledges support from the Science and Technology Facilities Council (STFC) through grant ST/Y000951/1.  {FST gratefully acknowledges the hospitality of the Department of Astronomy and Space Sciences at Istanbul University during her visit, which supported the writing of this review chapter.}

\bibliographystyle{abbrvnat-maxbibnames4}
\bibliography{chapter} 

\end{document}

%% file: journal-names.tex
\newcommand{\actaa}{Acta Astron.} % Acta Astronomica
\newcommand{\araa}{ARA\&A} % Annual Review of Astron and Astrophys
\newcommand{\aar}{A\&ARv} % Astrononmy \& Astrophysics Review
\newcommand{\aapr}{A\&ARv} % Astronomy\&Astrophysics Reviews
\newcommand{\ab}{Astrobiol.} % Astrobiology
\newcommand{\aj}{AJ} % Astronomical Journal
\newcommand{\apj}{ApJ} % Astrophysical Journal
\newcommand{\apjl}{ApJL} % Astrophysical Journal, Letters
\newcommand{\apjs}{ApJSS} % Astrophysical Journal, Supplement
\newcommand{\ao}{Appl. Opt.} % Applied Optics
\newcommand{\apss}{Astro. \& Space Sci.} % Astrophysics and Space Science
\newcommand{\aap}{A\&A} % Astronomy and Astrophysics
\newcommand{\aaps}{A\&AS.} % Astronomy and Astrophysics, Supplement
\newcommand{\baas}{Bull. Am. Astron. Soc.} % Bulletin of the AAS
\newcommand{\caa}{Chinese A\&A} % Chinese Astronomy and Astrophysics
\newcommand{\cjaa}{Chinese J. A\&A} % Chinese Journal of Astronomy and Astrophysics
\newcommand{\cqg}{Class. Quantum Gravity} % Classical and Quantum Gravity
\newcommand{\gal}{Galaxies} % Galaxies
\newcommand{\gca}{Geo. Cosmo. Acta} % Geochimica Cosmochimica Acta
\newcommand{\icarus}{Icarus} % Icarus
\newcommand{\jcap}{JCAP} % Journal of Cosmology and Astroparticle Physics
\newcommand{\jgr}{J. Geophys. Res.} % Journal of Geophysics Research
\newcommand{\jgrp}{J. Geophys. Res. Planets} % Journal of Geophysics Research: Planets
\newcommand{\jqsrt}{J. Quant. Spectrosc. Radiat. Transf.} % Journal of Quantitiative Spectroscopy and Radiative Transfer
\newcommand{\memsai}{Mem. SAIt} % Mem. Societa Astronomica Italiana
\newcommand{\mnras}{MNRAS} % Monthly Notices of the RAS
\newcommand{\nat}{Nature} % Nature
\newcommand{\nastro}{Nat. Astron.} % Nature Astronomy
\newcommand{\ncomms}{Nat. Commun.} % Nature Communications
\newcommand{\nphys}{Nat. Phys.} % Nature Physics
\newcommand{\na}{New Astron.} % New Astronomy
\newcommand{\nar}{New Astron. Rev.} % New Astronomy Review
\newcommand{\physrep}{Phys. Rep.} % Physics Reports
\newcommand{\pra}{Phys. Rev. A} % Physical Review A: General Physics
\newcommand{\prb}{Phys. Rev. B} % Physical Review B: Solid State
\newcommand{\prc}{Phys. Rev. C} % Physical Review C
\newcommand{\prd}{Phys. Rev. D} % Physical Review D
\newcommand{\pre}{Phys. Rev. E} % Physical Review E
\newcommand{\prx}{Phys. Rev. X} % Physical Review X
\newcommand{\prl}{Phys. Rev. Let.} % Physical Review Letters
\newcommand{\psj}{Planet. Sci. J.} % Planetary Science Journal
\newcommand{\planss}{Planet. Space Sci.} % Planetary Space Science
\newcommand{\pnas}{Proc. Natl Acad. Sci. USA} % Proceedings of the US National Academy of Sciences
\newcommand{\procspie}{Proc. SPIE} % Proceedings of the SPIE
\newcommand{\pasa}{PASA} % Publications of the Astron.  Soc. of Australia
\newcommand{\pasj}{PASJ} % Publications of the Astron.  Soc. of Japan 
\newcommand{\pasp}{PASP} % Publications of the Astron.  Soc. of the Pacific
\newcommand{\rmxaa}{RMXAA} % Revista Mexicana de Astronomia y Astrofisica
\newcommand{\sci}{Science} % Science
\newcommand{\sciadv}{Sci. Adv.} % Science Advances
\newcommand{\solphys}{Sol. Phys.} % Solar Physics
\newcommand{\sovast}{Soviet Ast.} % Soviet Astronomy
\newcommand{\ssr}{Space Sci. Rev.} % Space Science Reviews
\newcommand{\uni}{Universe} % Universe